\documentstyle[12pt]{article}

\title{Simple description of
metastable phase decay on  wide
spectrum of activities}

\author{V.Kurasov}

\date{Victor.Kurasov@pobox.spbu.ru}

\begin{document}

\maketitle

One of the typical examples of the system evolution in the first order
phase transition is the metastable phase decay in the closed system.
After the instantaneous creation of initial supersaturation
$\Phi$ the external
influence doesn't take place and the evolution of
the supersaturation $\zeta$ in the system occurs only
due to the nucleation kinetics.
The supercritical embryos of a new liquid phase, i.e.  the droplets, appear
mainly on heterogeneous centers which inevitably exist in any real system.
The number of heterogeneous centers  slightly changes in time, but their
activation barriers can attain rather arbitrary values. Then one can
speak about the distribution $\eta_{tot}$ of the heterogeneous center numbers
over
the heights
$\Delta F$ of activation barriers (i.e over activities $w$ of heterogeneous
centers).

It is natural to suppose that the mentioned distribution is rather smooth
as a function of $w$
and covers some region of the activation barriers heights. The subject
of current publication is to give the most simple variant for the nucleation
description (the full theory is given in \cite{book}).
The property of activity spectrum to be the wide one will be seen a few
moments later.

The nucleation description has to take into account the processes of the
mother phase  consumption  and  the  processes   of  exhaustion  of
heterogeneous
centers. Then one comes to the non-linear system of condensation equations
where the behavior of $\zeta$ depended on $I$ in all previous moments
of time. The approximate method to solve this non-linear problem is
given below.

We consider the homogeneous system of a unit volume. The regime of the
droplets growth
is supposed to be the free molecular one, the process will be an isothermal
one.

One can prove analytically the validity of quasistationary approximation
for the nucleation rate  for  all  situations   when  the  critical
embryo has
at least few dozens of molecules.

Under these conditions one can also prove that the main role in vapor
consumption belongs to the supercritical embryos (which are called as
droplets).

The rate of nucleation $I$ is the intensity of
the droplets formation. To characterize the droplet with $\nu$ molecules
it is convenient to
introduce the linear size $\rho= \nu^{1/3}$. The rate of $\rho$ growth
is $\zeta / \tau$ with some characteristic constant $\tau$ for all sizes
of droplets.

One can easily extract the main dependencies in the nucleation rate
behavior
over the supersaturation $\zeta$  and over the free (unoccupied  by
droplets)
number $\eta$ of the heterogeneous
centers: $I \sim \exp(- \Delta F) \eta$, where $\Delta F$ is the height
of the activation barrier taken in thermal units.
The end of nucleation corresponds to the relatively small fall of
supersaturation
from  initial value $\Phi$. Then one can linearize $\Delta F$ as a
function of $\zeta$ at  the  nucleation  period  and  come  to  the
following
approximation
$$
I(\zeta,\eta) = I(\Phi, \eta_{tot}) ) (\eta / \eta_{tot})
\exp(\Gamma (\zeta - \Phi)/ \Phi)
$$
 where $\Gamma =  - \Phi  d \Delta F / d \zeta
|_{\zeta = \Phi}$.
 One can also see that during the nucleation the rate
of droplets growth $\zeta/ \tau$ is approximately constant in time $\Phi /
\tau$.

To get the behavior of the supersaturation one has to know $\zeta (t)$.
It can be gotten by the balance equation
$$
\zeta = \Phi - G/n_{\infty}
$$
where $G$
is the number of the molecules in the liquid phase and
 $n_{\infty}$ is the number of the molecules in the saturated
vapor. For $G$ one can get $G = \int dw g(w)$, where the integral is taken
over the whole spectrum of activities and $g(w)$ is the number of molecules
in the droplets on the centers with activity $w$. The last value is equal
to
$$
g(w) = \int_0^t (t-t')^3  (\frac{\Phi}{\tau})^3 I(t',w) dt'
$$

Having mentioned that the main dependence of $I$ over the activity $w$
is $\sim \exp(-w)$ one can come to
$$
G = \int dw \int_0^t (t-t')^3 \exp(-w+w_*)
\frac{\eta(w,t')}{ \eta_{tot} (w_*)}
(\frac{\Phi}{\tau})^3  I(w_*, \Phi, \eta_{tot}(w_*) )
\exp(\Gamma \frac{\zeta(t') - \Phi}{\Phi} ) dt'
$$
with some parameter $w_*$.
In the last equation there exists $\eta(w,t')$. It can be given by the
following expression
$$
\eta(w,t) = \eta_{tot} (w)
\exp(-\int_0^t  \exp(-w+w_*)
\frac{ I(w_*, \Phi, \eta_{tot}(w_*) )}{\eta_{tot} (w_*)}
\exp(\Gamma \frac{\zeta(t') - \Phi}{ \Phi} ) dt' )
$$
The last two equations form
the closed system of condensation equations.

One can see that these equations are essentially non-linear ones.

The choice of $w_*$ is rather arbitrary.
>From the last equation one can see that if $w_*$ is
chosen to satisfy condition $\eta(w_*, t= \infty) = \eta_{tot} (w_*)
/2$ then all centers with $w>w_*+1$ remain practically unexhausted
and all centers with $w<w_* - 1$ are practically exhausted. So, the
interesting
region $[w_* -1 , w_* + 1 ]$ is rather narrow. One can put in
$[w_* -1 , w_* +1]$
the value $\eta_{tot} (w) $ to $\eta_{tot} (w_*)$.

Distribution  $\eta_{tot} (w)$ has to be limited from below to ensure
the finite value of integral nucleation rate. Denote by $w_0$ the minimum
of $w$. If $w_* -w_0 \geq  (3 \div 4)$ then one can say that
the relatively "wide spectrum" of
activities takes place in nucleation.
In the opposite situation one can consider nucleation as taking place
on the similar heterogeneous centers.

Consider the situation of wide spectrum.
We need to know the behavior of supersaturation at the end of nucleation
when the fall of supersaturation stops the nucleation.
All centers with $w>w_+ \equiv w_*+2$ will be exhausted  until  the
first quarter
of
the nucleation period duration. The droplets formed on these centers can
be regarded as  formed at the initial moment of time. Then one can
write for $G$:
$$
G = \int_{-\infty}^{w_+} \eta_{tot} (w)  dw t^3
(\frac{\Phi}{\tau})^3
$$
$$
+
 \int_{w_+}^{\infty} dw \int_0^t (t-t')^3 \exp(-w+w_*)
\frac{\eta(w,t')}{ \eta_{tot} (w_*)}
(\frac{\Phi}{\tau})^3  I(w_*, \Phi, \eta_{tot}(w_*) )
\exp(\Gamma \frac{\zeta(t') - \Phi}{ \Phi} ) dt'
$$

The second term is negligible in comparison with the first one. Then
$$
G = \int_{-\infty}^{w_+} \eta_{tot} (w)  dw t^3
(\frac{\Phi}{\tau})^3
$$
and the behavior of supersaturation is determined.
Then
$$
\zeta = \Phi -
 \int_{-\infty}^{w_+} \eta_{tot} (w)  dw t^3
(\frac{\Phi}{\tau})^3 / n_{\infty}
$$

The last step to do is to determine $w_*$. For $w_*$ we have
$$
\frac{1}{2} =
\exp(  - \int_0^{\infty}
 \frac{ I(w_*, \Phi, \eta_{tot}(w_*) )}{\eta_{tot} (w_*)}
\exp(\Gamma \frac{\zeta(t') - \Phi}{ \Phi } ) dt'              )
$$
or
$$
\frac{1}{2} =
\exp(  - \int_0^{\infty}   \frac{ I(w_*, \Phi, \eta_{tot}(w_*) )}{\eta_{tot}
(w_*)}
\exp( - \frac{\Gamma}{\Phi n_{\infty}}
 \int_{-\infty}^{w_+} \eta_{tot} (w)  dw t'^3
(\frac{\Phi}{\tau})^3
  ) dt'  )
$$
Having calculated the integral one can come to
$$
\frac{1}{2} =
\exp( -   \frac{ I(w_*, \Phi, \eta_{tot}(w_*) )}{\eta_{tot} (w_*)}
[ \frac{\Gamma}{ n_{\infty} \Phi }
 \int_{-\infty}^{w_+} \eta_{tot} (w)  dw
(\frac{\Phi}{\tau})^3  ]^{-1/3}  0.9 )
$$
The last equation together with an
explicit expression for $I$ given by the classical theory of nucleation is
an ordinary algebraic equation.

The number $N_{\infty} (w)$ of the droplets formed on the centers with
activity $w$ can be calculated as
$$
N_{\infty} (w) =
\eta_{tot} (w) (1-
\exp( - \exp(-w+w_*)  \frac{ I(w_*, \Phi, \eta_{tot}(w_*) )}{\eta_{tot} (w_*)}
$$
$$
[ \frac{\Gamma}{\Phi n_{\infty}}
 \int_{-\infty}^{w_+} \eta_{tot} (w)  dw
 ]^{-1/3}
\frac{\tau}{\Phi}  0.9 ) )
$$

The total number of droplets can be calculated by the integration over
all $w$.

Consider the opposite situation. The spectrum of activities  participating
in the nucleation process is so narrow the one can consider all centers
as the similar ones. Then
the system of equations will be the following
$$
\eta(t) = \eta_{tot}
\exp ( -\frac{I(\Phi, \eta_{tot} )}{\eta_{tot}}
\int_0^t  \exp(-\frac{ \Gamma g(t')}{n_{\infty}  \Phi}  ) dt' )
$$
$$
g(t) = I(\Phi,\eta_{tot})
\int_0^t (t-t')^3  \frac{\Phi^3}{\tau^3}
  \exp(-\frac{ \Gamma g(t')}{n_{\infty}  \Phi}  )
\frac{\eta}{\eta_{tot}} dt'
$$

One can use iterations defined as $\eta_{i+1} =  Q(g_i)$,
$g_{i+1} = P(g_i, \eta_i)$ with
initial approximations $g_0 = 0$, $\eta_0 = \eta_{tot}$. Here
$Q$, $P$ define the r.h.s. of two previous equations.
The second iteration gives already suitable results.
Then one can  get with a relatively high precision
for the total number of droplets
$$
N(\infty)  =  \eta_{tot}
[1-
\exp[ -
0.9
\frac{I(\Phi, \eta_{tot})^{3/4} }{\eta_{tot}}
(
\frac{4 n_{\infty} \tau^3}{\Gamma \Phi^2})^{1/4} ]]
$$
The last relation solves the problem.

\end{document}